\documentclass[11pt]{article}
\usepackage{graphicx,amsmath,amssymb,epsf} \usepackage{latexsym,bm,
  slashed} \usepackage{xcolor} \definecolor{dark}{rgb}{0.10,0.2,0.3}
\definecolor{magenta}{rgb}{0.7,0.1,0.3}
\definecolor{purpure}{rgb}{0.5,0.15,0.3}
\usepackage[font=small,format=plain,labelfont=bf,up,textfont=it,up]{caption}
\usepackage{hyperref, cite} \hypersetup{colorlinks, citecolor=blue,
  filecolor=blue, linkcolor=magenta,
  urlcolor=purpure,hyperfootnotes=true,pdftex} 
\newcommand{\tr}{{\rm tr}}
\newcommand{\kt}{{\bm k}}

\newcommand{\qt}{{\bm q}}
\newcommand{\xt}{{\bm x}}
\newcommand{\rt}{{\bm r}}

\newcommand{\zt}{{\bm z}}
\newcommand{\rht}{{\bm \rho}}
\newcommand{\bt}{{\bm b}}
\newcommand{\yt}{{\bm y}}
\newcommand{\asb}{\overline{\alpha}_s}
\newcommand{\asbt}{\tilde{\alpha}_s}

 \title{\bf The  Pomeron loop as a perturbative correction to single Pomeron exchange revisited}
 \author{Martin Hentschinski, Karina Mendoza-Ram\'irez, Miguel A.  Oca\~na-Bribiesca
 \\ \\
Departamento de Actuar\'ia, F\'isica y Matem\'aticas,
Universidad de las Am\'ericas Puebla, \\ Santa Catarina M\'artir, 72820 Puebla, Mexico }

\begin{document}

\maketitle

\begin{abstract}
 Pomeron loops are known to arise naturally as building blocks of  high energy scattering amplitudes as soon as one starts to consider corrections to single Pomeron exchange, i.e. to  high energy resummation based on the Balitsky-Fadin-Kuraev-Lipatov evolution equation. While some authors argue that Pomeron loops provide only very small corrections, a 2013 study found that the complete QCD Pomeron loop provides a large correction, which can exceed the single Pomeron contribution already at LHC energies. In this paper we carefully review the derivation of the single Pomeron loop and evaluate  the resulting expression  numerically, making use of a representation of the triple Pomeron vertex by Korchemsky. While we do not confirm the findings of the 2013 study, we find that the Pomeron loop can provide a $24\%-39\%$ correction at lowest currently accessible values of $x$ and hard scales at the non-perturbative boundary. 
\end{abstract}

\section{Introduction}

The high energy limit of strongly interacting cross-sections is non-trivial even in the presence of a hard scale $M$ which renders the strong coupling constant small  $\alpha_s(M) \ll 1$. With  $x = M^2/s$, the  perturbative Regge
limit of Quantum Chromodynamics (QCD) is provided through the limit  $x\to 0$, which corresponds to the limit $\sqrt{s}$ at a fixed hard scale $M^2$. At small, but still  moderate values of $x$, the QCD high energy limit is provided by the  Balitsky-Fadin-Kuraev-Lipatov (BFKL) evolution
equation \cite{Kuraev:1977fs,Kuraev:1976ge, Balitsky:1978ic} which
resums perturbative terms $\left(\alpha_s \ln 1/x \right)^n$ to all
orders in $\alpha_s$. It gives rise to the effective $t$-channel exchange of the BFKL Pomeron, which is obtained as an interacting two reggeized gluon exchange, where a reggeized gluon denotes an effective high energy gluon degree of freedom. Since the BFKL Pomeron grows like a power  $\sim x^{-\lambda}$, at some value of $x$, the perturbative suppression of multiple reggeized gluon exchange through the smallness of the coupling to external particles is compensated and must be taken into account. A framework which achieves this is provided by the JIMWLK-BK evolution equations\footnote{ Jalilian-Marian-Iancu-McLerran-Weigert-Leonidov-Kovner (JIMWLK), Balitsky-Kovchegov (BK)}  \cite{Balitsky:1995ub,
  Jalilian-Marian:1997ubg, Kovchegov:1999yj, Iancu:2000hn,
  Weigert:2000gi, Iancu:2001ad, Ferreiro:2001qy}, which resum such contributions for a dilute-dense scenario, i.e. scattering of a dilute projectile  on a dense target. Unlike the BFKL equation, this framework takes into account an arbitrary number of (reggeized) gluon exchanges in the formulation of low $x$ evolution equations, as long as they couple to the dense target. Within a multi-color $N_c \to \infty$ approximation, as provided by the BK equation, it is again possible to use the Pomeron as an effective degree of freedom in the $t$-channel, which  somehow simplifies the discussion. In addition to BFKL dynamics, one then encounters the triple-Pomeron vertex  \cite{Bartels:1994jj, Bartels:1995kf}, which describes the splitting of a single Pomeron into two; the resulting evolution equation then resums so-called fan diagrams, where the number of Pomeron increases in each evolution step in $\ln 1/x$, when moving from the dilute projectile towards the dense target. 

While phenomenologically successful, it is well known that this framework does not include fluctuations in the number of exchanged Pomerons, so-called Pomeron loops. The emergence  of Pomeron loops has been first realized within pre-QCD Reggeon Field Theory  \cite{Gribov:1968nmj,Gribov:1968uy,Abarbanel:1973pq,Sugar:1974td, Bartels:2015gou,Bartels:2024jfb}, with a non-perturbative Pomeron as central degree of freedom and it  has  later on also been explored within the BK-JIMWLK setup, see e.g. \cite{Iancu:2004iy,Blaizot:2005vf,Kovner:2005nq,Dumitru:2007ew} and  within a  framework based on  reggeized gluon degrees of freedom, see e.g. \cite{Bartels:2002au,Braun:2009sh,Flensburg:2010kq,Levin:2025gbh,Bartels:2007dm,Lipatov:1995pn} as well as \cite{Mueller:1996te,Salam:1995uy,Rembiesa:2005gj}. Indeed, as soon as splitting of a single Pomeron into two Pomerons is included in the description, Pomeron loops arise automatically;  the only exceptions are scattering of a dilute projectile on a dense target, since the large number of color sources in the latter enhances the splitting configuration \cite{Kovchegov:1999yj}. 

The numerical size of the Pomeron loop is therefore needed for various reasons: first, for the search of gluon saturation and the on-set of non-linear QCD evolution, e.g. \cite{Hentschinski:2022xnd,Aguilar:2024otb,Peredo:2023oym,Hentschinski:2025ovo,Mantysaari:2025ltq,Bautista:2016xnp,ArroyoGarcia:2019cfl,Hentschinski:2020yfm} which is commonly based on solutions of the BK equation, which does not contain Pomeron loop effects. It has also been argued that Pomeron loop effects are relevant for the determination of color dipole multiplicities \cite{Contreras:2025zsc,Contreras:2026gvm,Kutak:2025tsx,Kutak:2025syp,Kovner:2020exf}, which are of interest for studies of (entanglement) entropy in low $x$ Deep Inelastic Scattering, see e.g. \cite{Hentschinski:2023izh,Hentschinski:2022rsa,Hentschinski:2024gaa,Hentschinski:2026otq}.

Through the study  of a (1+1) dimensional model, which mimics the QCD evolution including Pomeron loops,  it has been argued in  \cite{Dumitru:2007ew} that fluctuations in the number of Pomerons are strongly suppressed by running coupling effects. This argument is however in sharp contrast to the findings of  \cite{Braun:2013tha} who considered a single Pomeron loop as a perturbative correction to single Pomeron exchange. The authors of this study found that the Pomeron loop correction dominates single Pomeron exchange at values of $x$ which are currently within reach of LHC experiments. Since  \cite{Braun:2013tha} consider the Pomeron loop within a complete (3+1) dimensional setup, this suggests that the Pomeron loop configuration might be after all not as small as thought and that it needs to be considered for a  complete phenomenology of the QCD high energy limit.

In the following we will re-investigate carefully the results of \cite{Braun:2013tha} and slightly extend them to the scattering of two virtual photons, with different photon virtualities. In this way we will not only study the mere size of the Pomeron loop correction, but we will have also access to the collinear limit, where one photon is significantly smaller than the other. We further aim at a further clarification of the theory expressions which underlies the study of  \cite{Braun:2013tha}. While as  \cite{Braun:2013tha} we will rely on results existing in the literature, such as the solution of the BFKL equation for finite momentum transfer by Lipatov,\cite{Lipatov:1985uk}, as well as the projection of the triple Pomeron vertex on conformal eigenfunctions by Korchemsky,  \cite{Korchemsky:1997fy}, we attempt a complete rederivation of the single Pomeron loop result. To this end, we will use that  QCD in the high energy limit  can be formulated as an effective field theory of reggeized gluons, where vertices and evolution kernels can be obtained from the QCD high energy effective action,  see \cite{Lipatov:1995pn,Hentschinski:2020rfx, Hentschinski:2011xg,Nefedov:2019mrg,Hentschinski:2021lsh}. While the BFKL Pomeron and triple Pomeron vertex arise naturally within this framework, it is also well known how to obtain within this setup the well-known JIMWLK-BK equation, see \cite{Hentschinski:2018rrf}.

The outline of this paper is as follows: In Sec.~\ref{sec:pomeron} we relate different approaches to the single Pomeron exchange and set our notation for the exploration of the Pomeron loop, which we explore in full detail in Sec.~\ref{sec:pomeron_loop} while Sec.~\ref{sec:pheno} provides  our numerical results. In Sec.~\ref{sec:concl} we provide our conclusions.

\section{The Single Pomeron Exchange  Contribution}
\label{sec:pomeron}
To have a good control over the relative weight of different contributions, we present the following discussion for the scattering of two color dipoles. Since we work always in the dilute approximation, this does not provide any substantial limitation of our result. It comes with the advantage that it allows for a straightforward connection to other results present in the literature. 

\subsection{Single Pomeron exchange for dipole-dipole scattering}
\label{sec:color_dipole}

We start with  the dipole amplitude defined in terms of reggeized gluon fields,
\begin{align}
  \label{eq:color_dipole}
  \hat{N}_\pm( \zt_1, \zt_2) & \equiv  \frac{1}{N_c} \tr \left[1\!\!1-V[\alpha_\pm](\zt_1)V[\alpha_\pm]^\dagger(\zt_2) \right],
\end{align}
where  we use the effective action framework of \cite{Lipatov:1995pn} to have a precise matching to reggeized gluon fields, following \cite{Hentschinski:2018rrf}. In particular 
\begin{align}
  \label{eq:Wilsonline}
  V[\alpha_\pm](\zt) & = \exp \left(ig \alpha_\pm^c t^c \right),
&
 V^\dagger[\alpha_\pm](\zt) & = \exp \left(-ig \alpha_\pm^c t^c \right)
\end{align}
with SU($N_c$) generators normalized to $\tr(t^a t^b) = \delta^{ab}/2$ and $3+1$ dimensional reggeized gluons fields,
\begin{align}
  \label{eq:reggeized_fields1}
  A_\pm(x) & = -i2 t^a\alpha^a_\pm (\xt) \delta(x^\pm), & \partial_\mp  A_\pm(x) & =0,
\end{align}
parameterized in terms of two dimensional fields $ \alpha_\pm (\xt)$ with 
\begin{align}
  \label{eq:reggeized_fields}
   \left  \langle  \alpha^a_+(\xt) \alpha^b_-(\yt)  \right\rangle & = i\delta^{ab}\int \frac{d^2 \qt}{(2 \pi)^2}  \frac{e^{i \qt \cdot (\xt- \yt)}}{\qt^2} = -i \delta^{ab} \frac{\ln |\xt- \yt|}{2 \pi},
\end{align}
where $A_\pm(x)$ denote reggeized gluon fields of this high energy effective action, where we refer to \cite{Hentschinski:2018rrf} for details as well as .  To leading order in the gauge coupling we have
\begin{align}
  \label{eq:N_expand}
  N_\pm(\xt_1, \xt_2) & = \frac{g^2}{2N_c} \left[\alpha_\pm^a(\xt_1)\alpha_\pm^a(\xt_2) - \frac{(\alpha_\pm^a(\xt_1))^2}{2}  - \frac{(\alpha_\pm^a(\xt_2))^2}{2} \right].
\end{align}
To describe the scattering of two color dipoles (which are assumed to be moving in opposite directions of the light-cone), we require the correlator $\langle N_+ N_-\rangle$ for which we find at tree-level
\begin{align}
  & \left\langle N_+ \left(\bt + \frac{\rt}{2},\bt - \frac{\rt}{2}  \right) 
  N_- \left(\bt' + \frac{\rt'}{2},\bt' - \frac{\rt'}{2}  \right)\right\rangle_{\rm tree} 
 = 
 \frac{\alpha_s^2(N_c^2-1)}{(4\pi)^2N_c^2} \int \frac{d^2 \kt \, d^2 \qt}{\kt^2 (\qt - \kt)^2}   e^{i \qt \cdot \left(\bt + \frac{\rt}{2} \right) }  \notag \\
& \hspace{1cm}
 e^{i \qt \cdot \left(\bt' + \frac{\rt'}{2} \right) } \left(1- e^{i \rt \cdot \left(\kt - \frac{\qt}{2} \right)} \right)
\left(1- e^{-i \rt \cdot \left(\kt + \frac{\qt}{2} \right)} \right)
\left(1- e^{-i \rt' \cdot \left(\kt - \frac{\qt}{2} \right)} \right) 
\left(1- e^{-i \rt' \cdot \left(\kt + \frac{\qt}{2} \right)} \right).
\end{align}
The relation of the dipole-dipole correlator to a physical scattering amplitude has been established in \cite{Bartels:2003yj}, which we use in the following to fix the overall normalization. With the relation between the Bartels-Golec Biernat-Peters (BGP) dipole-dipole amplitude and the above correlator given by 
\begin{align}
  \label{eq:BGP_relation}
  \pi^2 \delta^{(2)}(\qt - \qt') N_{\rm BGP}\left(\rt, \rt', \frac{\qt}{2} \right) & =
e^{-i \rt \cdot \frac{\qt}{2}}e^{i \rt' \cdot \frac{\qt'}{2}} \int d^2 \bt   e^{-i \qt \cdot \bt} 
\int d^2 \bt' e^{-i \qt' \cdot \bt'}  \notag \\
& \hspace{1cm}
\left\langle N_+ \left(\bt + \frac{\rt}{2},\bt - \frac{\rt}{2}  \right) 
  N_- \left(\bt' + \frac{\rt'}{2},\bt' - \frac{\rt'}{2}  \right)\right\rangle_{\rm tree}, 
\end{align}
we have for the elastic  scattering amplitude  of two hadrons $A,B \to A'B'$, 
\begin{align}
  \label{eq:A_v2}
  \delta^{(2)}(\qt - \qt')&  \mathcal{A}(s,\qt^2)  =  \frac{is}{\pi^2} \int d^2 \rt \int d^2 \bt \int_0^1 \frac{dz}{4 \pi} (\psi_A^* \psi_{A'})(\rt, z)
 e^{-i[\bt - \frac{(1-2z)}{2}\rt]\cdot \qt} \notag \\
& \hspace{-.5cm}\cdot 
\int d^2 \rt' \int d^2 \bt' \int_0^1 \frac{dz'}{4 \pi} (\psi_B^* \psi_{B'})(\rt', z') e^{i[\bt' - \frac{(1-2z')}{2}\rt]\cdot \qt'} \left \langle  N_+(\xt_1, \xt_2)  N_-(\xt'_1, \xt'_2) \right \rangle,
\end{align}
where $x_{1,2} = \bt \pm \rt/2$, $x_{1,2}' = \bt' \pm \rt'/2$ as usually and $\psi_i$, $i = A, A', B. B'$ are  light-front wave functions of external scattering particles.   Eqs.~\eqref{eq:BGP_relation}~\eqref{eq:A_v2} provides then a precise matching between dipole correlator and scattering amplitude, at the very least within a leading logarithmic approximation.

\subsection{The conformal solution to the non forward BFKL equation}
\label{sec:nonfwdBFKL}
With the relation of scattering amplitude and dipole dipole correlator established at tree-level, we would like now to resum high energy logarithms through making use of the conformal solution to the non-forward BFKL Pomeron within a leading logarithmic approximation by Lipatov  \cite{Lipatov:1985uk}.
To set the notation we first recall important  details on the formulation of this solution.  For an elastic scattering amplitude with a colorless $t$ changes  we have \cite{Lipatov:1985uk}, see also  \cite{Chachamis:2022jis,Lipatov:1996ts},
\begin{align}
  \label{eq:elastic_amplitude}
  \mathcal{A} (s,t) & = is \int \limits_{\delta-i\infty}^{\delta + i \infty} \frac{d\omega}{2 \pi i} e^{\omega Y} F_\omega(\qt^2), \qquad t = - \qt^2, 
\end{align}
where the contour of integration runs to the right of the singularities of the  partial wave amplitude $F_\omega$, which can be written as the following convolution,
\begin{align}
  \label{eq:partial_wave}
  F_\omega(\qt^2) & = \int d^2 \kt d^2 \kt'\Phi_1(\kt, \qt)\Phi_2(\kt', \qt) f_\omega(\kt, \kt', \qt).
\end{align}
Here $f_\omega$ denotes the Mellin transform of the non-forward BFKL Green's function to be specified below, while $\Phi_{1,2}$ are impact factors which describe the coupling of the Green's function to external scattering particles. They obey  the property
\begin{align}
  \label{eq:impact_factors}
  \Phi_{1,2}(0, \qt) & =  \Phi_{1,2}(\qt, \qt) = 0,
\end{align}
which can be related to gauge invariance of the hadronic impact factors. Expressing the Green's function through its Fourier transform, 
\begin{align}
  \label{eq:eq3_lev}
  \delta^{(2)}(\qt - \qt') f_\omega(\kt, \kt', \qt) & = \frac{1}{(2 \pi)^8} \prod_{r=1}^2 d^2 \rt_r \prod_{r=1}^2 d^2 \rt_r' e^{i \kt \cdot \rt_1 + i (\qt - \kt)\cdot \rt_2 - i \kt' \cdot \rt_1' + i (\qt' - \kt')\cdot \rt_2' } \notag \\
& \hspace{4cm}  f_\omega(\rt_1, \rt_2; \rt_1', \rt_2'),
\end{align}
one has
\begin{align}
  \label{eq:Fomega}
\delta^{(2)}(\qt - \qt')  F_\omega(\qt^2) & = \prod_{r=1}^2 \int \frac{d^2 \rt_r}{(2\pi)^2}  \int \frac{d^2 \rt_r'}{(2\pi)^2} \Phi_1(\rt_1, \rt_2; \qt)\Phi_2(\rt_1', \rt_2', \qt') f_\omega(\rt_1, \rt_2; \rt_1', \rt_2') \notag \\
\Phi_1(\rt_1, \rt_2; \qt) & = \int d^2 \kt \Phi_{1,2}(\kt, \qt) e^{i \kt \cdot \rt_1}e^{i (\qt - \kt) \cdot \rt_2},
\end{align}
with 
\begin{align}
  \label{eq:condition_coord}
  \int d^2 \rt_1 \Phi_1(\rt_1, \rt_2; \qt) & =  \int d^2 \rt_2 \Phi_1(\rt_1, \rt_2; \qt)  = 0.
\end{align}
The above property  allows to obtain the solution to the non-forward BFKL equation as 
\begin{align}
  \label{eq:f_Lev}
  f_\omega(\rt_1, \rt_2; \rt_1', \rt_2') & = \sum_{n=-\infty}^\infty \int \limits_{-\infty}^\infty d\nu \int d^2 \rt_0  \frac{\nu^2 + \frac{n^2}{4}}{\left(\nu^2 + \frac{(n+1)^2}{4} \right) \left(\nu^2 + \frac{(n-1)^2}{4} \right)} \frac{1}{\omega - \omega(n, \nu)} \notag \\
&  E^{n,\nu}(\rt_1-\rt_0, \rt_2-\rt_0) E^{n,\nu*}(\rt'_1-\rt_0, \rt'_2-\rt_0).
\end{align}
Here
\begin{align}
  \label{eq:omega}
  \omega(n, \nu) & =  \asb \chi_0(n, \nu), \qquad \asb  = \frac{\alpha_s N_c}{\pi} \notag \\
 \chi_0( n, \nu) & = 2 \Psi(1) - \Psi\left(\frac{1+|n|}{2} + i \nu\right) - \Psi\left(\frac{1+|n|}{2} -i\nu\right).
\end{align}
For  $n = \pm 1$ the integral over $\nu$ needs to be interpreted as a principal value for, see \cite{Lipatov:1985uk} for details; we will not make use of this in the following. Finally
\begin{align}
  \label{eq:E}
   E^{n,\nu}(\rt_1-\rt_0, \rt_2-\rt_0) & = \left(\frac{r_{12}}{r_{10} r_{20}} \right)^{\frac{1-n}{2} + i \nu} 
 \left(\frac{r_{12}^*}{r_{10}^*  r_{20}^* } \right)^{\frac{1+n}{2} + i \nu},
\end{align}
with
\begin{align}
  \label{eq:complex_coord}
  r_{ij} & = r_i - r_j, & r_j & = x_j + i y_j = |\rt_i| e^{i\phi_j}.
\end{align}
To use these results for a determination of the Pomeron loop configuration, it is useful to have also the corresponding leading order expressions  ($\asb = 0)$ for the Green's function. One finds  in momentum space
\begin{align}
  \label{eq:f_mom}
  \delta^{(2)}(\qt - \qt')f_\omega(\kt, \kt', \qt)\bigg|_{\asb = 0} & = \frac{1}{\omega} \frac{1}{\kt^2 (\qt - \kt)^2} \delta^{(2)}(\kt'- \kt) \delta^{(2)}(\qt'- \qt) ,
\end{align}
while
\begin{align}
  \label{eq:f0_Lev}
  f_\omega^{(0)}(\yt_1, \yt_2;\yt_{1'}, \yt_{2'}) & = \frac{(2 \pi)^2}{\omega}
\ln |\yt_{11'}| \ln |\yt_{22'}|,
\end{align}
which can be r-written as
\begin{align}
  \label{eq:f0_Levmod}
  f_\omega^{(0)}(\yt_1, \yt_2;\yt_{1'}, \yt_{2'}) & = \frac{2 \pi^2}{\omega}
\ln  \frac{\ln |\yt_{11'}| \ln |\yt_{22'}|}{\ln |\yt_{12'}| \ln |\yt_{1'2}|} \ln  \frac{\ln |\yt_{11'}| \ln |\yt_{22'}|}{\ln |\yt_{12}| \ln |\yt_{1'2'}|},
\end{align}
using the properties Eq.~\eqref{eq:intdy1} of the impact factors  \cite{Lipatov:1985uk},  since the difference is only due to terms independent of one of the coordinates.

\subsection{Dipole correlators and BFKL evolution}
\label{sec:impact}

In the following we want to combine the frameworks of the last two paragraphs. First note that the leading order result  Eq.~\eqref{eq:N_expand} can be written as
\begin{align}
  \label{eq:impa_N}
   N_\pm(\xt_2, \xt_1) & = \int d^2 \yt_1 \int d^2 \yt_2   \alpha^a(\yt_1)\alpha^a(\yt_2) \Phi_N(\yt_1, \yt_2; \xt_1, \xt_2), \notag \\
\end{align}
with
\begin{align}
  \label{eq:impaN2}
  \Phi_N(\yt_1, \yt_2; \xt_1, \xt_2) & = 
\frac{g^2}{4 N_c}  \bigg[ 
\left( \delta^{(2)} (\xt_{1} - \yt_{1})  -   \delta^{(2)} (\xt_2- \yt_{1}) \right) \notag \\
& \hspace{4cm} \cdot \left( \delta^{(2)} (\xt_{2} - \yt_{2}) -  \delta^{(2)} (\xt_{1} - \yt_{2})   \right)
\bigg],
\end{align}
 which satisfies Eq.~\eqref{eq:condition_coord}, i.e.
\begin{align}
  \label{eq:intdy1}
  \int d^2 \yt_1 \Phi_N(\yt_1, \yt_2; \xt_1, \xt_2) &  =  0 =  \int d^2 \yt_2 \Phi_N(\yt_1, \yt_2; \xt_1, \xt_2). 
\end{align}
To evaluate the correlator of two color dipoles to leading order within the BFKL framework, we associate the  `$+$' and `$- $' dipoles to scattering particles with significantly different  rapidity $Y = y_A - y_B > 0$. Including the constraint $Y > 0$ through a Mellin integral, we have  with $\delta >0$
\begin{align}
  \label{eq:2dipoles}
   \int\limits_{\delta - i \infty}^{\delta + i \infty} \frac{d \omega}{2 \pi i } &e^{\omega Y} \frac{1}{\omega} \langle N_+(\xt_1, \xt_2)N_-(\xt_{1'}, \xt_{2'}) \rangle  =  -\frac{ (N_c^2 -1)}{(2 \pi)^4}  \prod_{i=1,2} \int d^2\yt_i  d^2\yt_{i'} \notag 
\\
&  \Phi_N(\yt_1, \yt_2; \xt_1, \xt_2)
\Phi_N(\yt_{1'}, \yt_{2'}; \xt_{1'}, \xt_{2'}) f_\omega^{(0)} (\yt_1, \yt_2;\yt_{1'}, \yt_{2'}),
\end{align}
 with $f_\omega^{(0)}$ given by Eq.~\eqref{eq:f0_Levmod}. Since high energy evolution of the color dipole in the dilute approximation provides the BFKL equation, $f_\omega^{(0)}$ turns after resummation  into the complete BFKL Green's function $f_\omega(\yt_1, \yt_2;\yt_{1'}, \yt_{2'})$, given in  Eq.~\eqref{eq:f_Lev}. 
Since this function has the property  $f_\omega(\yt_1, \yt_1;\yt_{1'}, \yt_{2'}) = 0 =  f_\omega(\yt_1, \yt_2;\yt_{1'}, \yt_{1'})$,  it is now  sufficient to consider only those terms in the impact factor  with different coordinates.  This means that in the integrals over $\yt_i$, only those combinations which yield $\yt_1 = \xt_1$,   $\yt_2 = \xt_2$ and $\yt_1 = \xt_2$,   $\yt_2 = \xt_1$ give a non-zero contribution. We further note that
\begin{align}
  \label{eq:E_relation}
  E(\yt_{20}, \yt_{10}) & = (-1)^n E(\yt_{10}, \yt_{20}),
\end{align}
and
\begin{align}
  \label{eq:E_relation2}
  \frac{1}{2} \left[E(\yt_{10}, \yt_{20})  +  E(\yt_{20}, \yt_{10}) \right] & =  E(\yt_{10}, \yt_{20})  \frac{1 + (-1)^n}{2}.
\end{align}
This means that for this particular kind of impact factor, we can restrict to even values of the conformal spin $n$ in the BFKL Green's function. Using the large $N_c$ approximation, we obtain the resummed dipole-dipole scattering amplitude through the replacement 
\begin{align}
  \label{eq:replacement}
   \Theta(y_A-y_B) \langle N_+(\xt_1, \xt_2)N_-(\xt_{1'}, \xt_{2'}) \rangle
 & \to
 -  \frac{\bar{\alpha}_s^2}{4 N_c^2}\int\limits_{\delta - i \infty}^{\delta + i \infty} \frac{d\omega}{2 \pi i}  \bar{f}_\omega(\xt_1, \xt_2; \xt_{1'}, \xt_{2'}) e^{\omega Y}
\end{align}
where
\begin{align}
  \label{eq:fbar}
  \bar{f}_\omega(\xt_1, \xt_2; \xt_{1'}, \xt_{2'}) & = \sum_{n = \text{even}} \int\limits_{-\infty}^\infty d\nu \int d^2 \xt_0 
 \frac{\nu^2 + \frac{n^2}{4}}{\left(\nu^2 + \frac{(n+1)^2}{4} \right) \left(\nu^2 + \frac{(n-1)^2}{4} \right)} \notag \\
& \hspace{2cm} \frac{1}{\omega - \omega(n, \nu)}  
E^{n,\nu}(\xt_{10}, \xt_{20})E^{n,\nu,*}(\xt_{1'0}, \xt_{2'0}).
\end{align}
 The expression will  prove useful in constructing the Pomeron loop in the following.

While the above arguments are sufficient to determine the dipole-dipole correlator within the leading logarithmic approximation to all orders in the strong coupling, it is also instructive to obtain the same result directly from the BK/BFKL equation.  The BK-JIMWLK equation for a single dipole reads
\begin{align}
  \label{eq:bk}
\pm \frac{d}{dy} \hat{N}_\pm (y;\xt_1, \xt_2) &  =  \frac{\asb}{2 \pi}\bigg[ \int d^2 \yt_3 \frac{(\xt_1 - \xt_2)^2}{(\xt_1 - \yt_3)^2 (\xt_2  - \yt_3)^2}
 \bigg( 
 \hat{N}_\pm (\xt_1, \yt_3)  +  \hat{N}_\pm (\yt_3, \xt_2) \notag \\
&  
\hspace{4cm}
-  \hat{N}_\pm (\xt_1, \xt_2)   -  \hat{N}_\pm (\xt_1, \zt)  \hat{N}_\pm (\zt, \xt_2) 
\bigg)
 \bigg],
\end{align}
where the `$\pm$' takes into account that for the minus side, evolution takes place towards negative rapidities. In the following, we treat the  quadratic term in $\hat{N}_\pm$ as  a perturbation, i.e. we take  $\hat{N}_\pm = \mathcal{O}(\alpha_s)$ and split $\hat{N}_\pm = \hat{N}_\pm^{(0)} +  \hat{N}_\pm^{(1)}$, where $\hat{N}_\pm^{(k)} = \mathcal{O}(\alpha_s^{1+k})$. For the leading term, we arrive at the BFKL equation for the dipole operator, 
\begin{align}
  \label{eq:bfkl_bk}
\pm \frac{d}{dy} \hat{N}^{(0)}_\pm (\xt_1, \xt_2) &  =  \frac{\asb}{2 \pi}\bigg[ \int d^2 \yt_3 \frac{(\xt_1 - \xt_2)^2}{(\xt_1 - \yt_3)^2 (\xt_2  - \yt_3)^2}
 \bigg( 
 \hat{N}_\pm^{(0)} (\xt_1, \yt_3)   \notag \\
& \hspace{5cm} +  \hat{N}_\pm^{(0)} (\yt_3, \xt_2)  -  \hat{N}_\pm^{(0)} (\xt_1, \xt_2)
\bigg)
 \bigg].
\end{align}
It is solved\footnote{See \cite{Lotter:1996vk} for an explicit construction} by the BFKL Green's function Eq.~\eqref{eq:f_Lev}, however with different initial conditions. We therefore  introduce at first a  version of the BFKL Green's function Eq.~\eqref{eq:f_Lev}  normalized to two  delta functions. Since
\begin{align}
  \label{eq:eq:f0_rel2}
  f^{(0)}_\omega (\xt_1, \xt_2; \yt_1,\yt_2)& =  
 \int d^2 \qt  \int d^2 \kt  e^{i \kt  (\xt_1 - \yt_1)}  e^{i (\qt- \kt)  (\xt_2 - \yt_2)} \frac{1}{\omega} \frac{1}{\kt^2 (\qt - \kt)^2};
\end{align}
and
\begin{align}
  \label{eq:Greensfun}
  \partial^2_{\yt_1} \partial^2_{\yt_2}f^{(0)}_\omega (\xt_1, \xt_2; \yt_1,\yt_2) & =  \frac{(2 \pi)^4}{\omega} \delta^{(2)}(\xt_1 - \yt_1)\delta^{(2)}(\xt_2 - \yt_2) 
\end{align}
we have with
\begin{align}
  \label{eq:Greensfun3}
  \tilde{g}(Y; \xt_1, \xt_2; \yt_1, \yt_2) & = \int \frac{d\omega}{2\pi i} e^{\omega Y} \frac{1}{(2\pi)^4} \partial^2_{\yt_1} \partial^2_{\yt_2}  f_\omega(\xt_1, \xt_2; \yt_1, \yt_2),
\end{align}
a solution to the BFKL equation which satisfies the necessary initial condition,
\begin{align}
  \label{eq:Greensfun4}
  \tilde{g}(Y=0; \xt_1, \xt_2; \yt_1, \yt_2)&=  \delta^{(2)}(\xt_1 - \yt_1)\delta^{(2)}(\xt_2 - \yt_2) .
\end{align}
A  solution to Eq.~\eqref{eq:bfkl_bk} is then given by 
\begin{align}
  \label{eq:N0}
  N_{\pm, res}^{(0)}( Y;\xt_1, \xt_2) & =  \prod_{i=1}^2  \int  d^2\yt_i \,\tilde{g}(\pm Y; \xt_1, \xt_2; \yt_1, \yt_2) \hat{N}( \yt_1, \yt_2).
\end{align}
%Even though we will use in the following this expression only in the approximation Eq.~\eqref{eq:N_expand}, it is valid in general 
%and we will use
%\begin{align}
%  \label{eq:Nopm}
%   N_+^{(0)}(y_A-y_B;\xt_1, \xt_2) & =  \prod_{i=1}^2  \int  d^2\yt_i \,\tilde{g}(y_A - y_B; \xt_1, \xt_2; \yt_1, \yt_2) \hat{N}( \yt_1, \yt_2), \notag \\
% N_+^{(0)}(0;\xt_1', \xt_2') & =\hat{N}( \xt_1', \xt_2').
%\end{align}
With
\begin{align}
  \label{eq:iM}
  i\mathcal{M}(y_A-y_B; \xt_1, \xt_2; \xt_1', \xt_2') & \equiv \left \langle N_+(y_A; \xt_1, \xt_2)N_+(y_B; \xt_1', \xt_2')  \right \rangle,
\end{align}
we have  
\begin{align}
  \label{eq:iM1w}
  \mathcal{M}_1(Y; \xt_1, \xt_2; \xt_1', \xt_2') & =  i \frac{\bar{\alpha}_s^2}{4N_c^2} \int\limits_{\delta - i \infty}^{\delta + i \infty} \frac{d\omega}{2 \pi i}  \sum_{n = \text{even}} e^{\omega Y} \int\limits_{-\infty}^\infty d\nu 
 \frac{\nu^2 + \frac{n^2}{4}}{\left(\nu^2 + \frac{(n+1)^2}{4} \right) \left(\nu^2 + \frac{(n-1)^2}{4} \right)} \notag \\
& \hspace{2cm} \frac{1}{\omega - \omega(n, \nu)}  
 \int d^2 \xt_0  E^{n,\nu}(\xt_{10}, \xt_{20})E^{n,\nu,*}(\xt_{1'0}, \xt_{2'0}),
\end{align}
where the subscript `$1$' indicates that we are dealing here with the exchange of a single Pomeron, in contrast to the Pomeron loop to be discussed in the next section.  We finally want to take the limit $\qt \to$ at the level of the physical scattering amplitude. First note that 
\begin{align}
  \label{eq:FTiM_1}
  \int d^2 \bt e^{-i\qt \cdot \bt}  \int d^2 \bt' e^{-i\qt' \cdot \bt'} i \mathcal{M}_1 & = -\delta^{(2)}(\qt - \qt') \frac{\asb^2 4 \pi^4}{N_c^2} \notag \\
&  \int d^2 \kt \int d^2 \kt' e^{-i \rt \cdot (\kt - \qt/2)}  e^{i \rt' \cdot (\kt' - \qt/2)} f(Y; \kt, \kt', \qt),
\end{align}
with the momentum space non-forward BFKL Green's function
\begin{align}
  \label{eq:nonfwdF}
  f(Y; \kt, \kt', \qt) & =
 \int\limits_{\delta - i \infty}^{\delta + i \infty} \frac{d\omega}{2 \pi i}  \sum_{n = \text{even}} e^{\omega Y} \int\limits_{-\infty}^\infty d\nu 
\frac{1}{\omega - \omega(n, \nu)} \notag \\
&  \frac{1}{2^6 \pi^4}  \int d^2 \rht \int d^2 \rht'
e^{i \rht \cdot (\kt - \qt/2)} e^{-i \rht' \cdot (\kt' - \qt/2)}
 \frac{ |\rht||\rht'| E_q^{n,\nu}(\rht) E_q^{n,\nu*}(\rht')}{\left(\nu^2 + \frac{(n+1)^2}{4} \right) \left(\nu^2 + \frac{(n-1)^2}{4} \right)},
\end{align}
where the functions $E_q^{n,\nu}(\rt)$ correspond to the so-called mixed representation\cite{Lipatov:1985uk}, i.e.
\begin{align}
  \label{eq:q_Efunc}
  \int d^2 \bt & 
e^{i \qt \cdot \bt} E^{n, \nu}(\bt + \rt/2, \bt - \rt/2)  =  \frac{b_{n,\nu}}{2\pi^2} {|\rt|}  E_q^{n, \nu}(\rt), 
%notag \\
% \int d^2 \bt
%e^{-i \qt \cdot \bt} E^{n, \nu*}(\bt + \rt/2, \bt - \rt/2) & =  \frac{b_{n,\nu}^*}{2\pi^2} {|\rt|}  E_q^{n, \nu*}(\rt),
\end{align}
with 
\begin{align}
  \label{eq:bnnu}
  \notag \\
b_{n\nu} & = \frac{\pi ^3 2^{4 i \nu } \Gamma \left(\frac{n+1}{2}-i \nu \right)
   \Gamma \left(\frac{n}{2}+i \nu \right)}{\left(\frac{n}{2}-i \nu
   \right) \Gamma \left(\frac{n}{2}-i \nu \right) \Gamma
   \left(\frac{n+1}{2}+i \nu \right)}, &  b_{n, \nu} b_{n\nu}^* & = \frac{4 \pi ^6}{n^2+4 \nu ^2}.
\end{align}
A result obtained by Chachamis and Sabio Vera demonstrated  in \cite{Chachamis:2022jis} finally  allows us to show that
\begin{align}
  \label{eq:limitqt0}
  \lim_{\qt \to 0}  \frac{1}{2^6 \pi^4}  \int d^2 \rht \int d^2 \rht' & 
e^{i \rht \cdot (\kt - \qt/2)} e^{-i \rht' \cdot (\kt' - \qt/2)}
 \frac{ |\rht||\rht'| E_q^{n,\nu}(\rht) E_q^{n,\nu*}(\rht')}{\left(\nu^2 + \frac{(n+1)^2}{4} \right) \left(\nu^2 + \frac{(n-1)^2}{4} \right)}  \notag \\ & = \frac{1}{2\pi^2} 
\frac{e^{i n (\theta_k - \theta_{k'})}}{(\kt^2)^{\frac{3}{2} - i \nu} (\kt^{'2})^{\frac{3}{2} + i \nu}}.
\end{align}
To have an explicit physical process, we  consider now the scattering of two virtual photons with transverse polarization with virtualities $Q_A^2$ and $Q_B^2$; the extension to longitudinal polarization and/or e.g. charmonium production is then straightforward.  Ignoring quark masses we have  \cite{Golec-Biernat:1998zce}
\begin{align}
  \label{eq:wf_photon}
  |\psi_T(z, \rt)|^2 & = \frac{2 N_c \alpha}{\pi} \sum_f e_f^2[z^2 + (1-z)^2]z(1-z)Q^2 K_1^2(\sqrt{z(1-z)}Q r).
\end{align}
Since the wave function depends only on the modulus $r = |\rt|$, we can make use of the integral 
\begin{align}
  \label{eq:intgamma}
  \int d^2 \kt \frac{1-e^{i \kt \cdot \rt}}{(\kt^2)^{\frac{3}{2} - i \nu}} & = \frac{\pi  2^{-1+2 i \nu } \Gamma \left(i \nu
   -\frac{1}{2}\right) (r^2)^{\frac{1}{2}-i \nu
   }}{\Gamma \left(\frac{3}{2}-i \nu \right)},
\end{align}
and with 
\begin{align}
  \label{eq:WF_Mellin}
  \int d^2 \rt \int_0^1 dz |\psi_T(z, \rt)|^2 \left(\rt^2 \right)^{\frac{1}{2} - i \nu} & = 
N_c H_T(\nu)
 \left(\frac{Q^2}{4} \right)^{-\frac{1}{2}+i \nu } \frac{ \Gamma \left(\frac{3}{2}-i \nu \right)}{-\Gamma \left(i \nu -\frac{1}{2}\right)},
\end{align}
where
\begin{align}
  \label{eq:HT}
  H_T(\nu) & = \sum_fe_f^2 \frac{ \pi ^2 \alpha  \left(4 \nu ^2+9\right) \tanh (\pi  \nu ) \text{sech}(\pi  \nu )
  }{16 \nu
   \left(1 + \nu ^2 \right)},
\end{align}
and noting that 
\begin{align}
  \label{eq:note}
  \frac{ \Gamma \left(\frac{3}{2}-i \nu \right)}{-\Gamma \left(i \nu -\frac{1}{2}\right)}\frac{ \Gamma \left(\frac{3}{2}+i \nu \right)}{-\Gamma \left(-i \nu -\frac{1}{2}\right)} = \left(\frac{1}{4} + \nu^2 \right)^2,
\end{align}
 we finally have from Eq.~\eqref{eq:A_v2}
\begin{align}
  \label{eq:scattering_cont3}
   \mathcal{A}_1(s,t=0) & =  i \frac{s}{Q_A Q_B} \cdot  \left(\frac{\asb^2}{8} \right) 
\int d\nu e^{Y \omega(0, \nu)} H_T^2(\nu)  \left(\frac{Q_A^2}{Q_B^2} \right)^{i \nu},
\end{align}
for the physical scattering amplitude of of two virtual photons with single Pomeron exchange.

\section{The Pomeron-loop}
\label{sec:pomeron_loop}

While the explicit expression for the triple Pomeron vertex is well known and has been used several times in the literature to formulate the Pomeron loop configuration \cite{Bartels:2007dm,Braun:2009sh}, the extraction  of the triple Pomeron vertex  based on multiple discontinuities is cumbersome and the combination into a Pomeron loop is subject to uncertainties as far as symmetry factors and Jacobian factors of high energy factorized amplitudes are concerned. We therefore attempt in the following an independent derivation which can be  verified relatively easily. To this end we start from the well known JIMWLK equation formulated for a single dipole, which is essentially the BK equation with the non-linear term not yet factorized, i.e. written for uncontracted reggeized gluon fields. Treating the non-linear term as a perturbation, we obtain in this way the BFKL Pomeron  and the triple Pomeron vertex with uncontracted reggeized gluon fields. Both the BFKL Pomeron and the Pomeron loop are  obtained through contracting the solution from the `$+$' and '$-$'sides at central rapidities.

\subsection{Determination of the Pomeron-loop configuration}

As first step we isolate the triple Pomeron vertex from the JIMWLK-BK equation. With $\hat{N}_\pm = \hat{N}_\pm^{(0)} +  \hat{N}_\pm^{(1)}$  we find 
\begin{align}
  \label{eq:bk_n1}
\pm\frac{d}{dy}& \hat{N}^{(1)}_\pm (y;\xt_1, \xt_2)   =  \frac{\asb}{2 \pi}\bigg[ \int d^2 \yt_3 \frac{(\xt_1 - \xt_2)^2}{(\xt_1 - \yt_3)^2 (\xt_2  - \yt_3)^2}
 \bigg( 
 \hat{N}^{(1)}_\pm (y;\xt_1, \yt_3)    \notag \\
&  \hspace{1cm}+  \hat{N}^{(1)}_\pm (y;\yt_3, \xt_2)
 -  \hat{N}^{(1)}_\pm (y;\xt_1, \xt_2) 
-  \hat{N}^{(0)}_\pm (y;\xt_1, \yt_3)  \hat{N}^{(0)}_\pm (y;\yt_3, \xt_2) 
\bigg)
 \bigg],
\end{align}
which is solved by 
\begin{align}
  \label{eq:N1}
  N_\pm^{(1)}(Y; \xt_1, \xt_2) & =\mp \frac{\asb}{2 \pi} \int^Y_{Y_0} dy 
\int \prod_{i=1}^2  d^2\rt_i
  \tilde{g}(\pm(Y-y); \xt_1, \xt_2; \yt_1, \yt_2)  \notag \\
&
\bigg[ \int d^2 \yt_3 \frac{(\yt_1 - \yt_2)^2}{(\yt_1 - \yt_3)^2 (\yt_2  - \yt_3)^2}
 \hat{N}^{(0)}_\pm (y;\yt_1, \yt_3)  \hat{N}^{(0)}_\pm (y;\yt_3, \yt_2).
\end{align}
We find a two Pomeron state which is evolved from rapidity $Y_0$ up to some rapidity $y$, where the two Pomerons merge into a single Pomeron. This Pomeron is finally  evolved up to the  rapidity $Y$. To construct from this the Pomeron loop we proceed as follows: we first divide our scattering amplitude into four rapidities, which we associate with the hadrons $A, A'$ and $B,B'$ as well as the two triple Pomeron vertices which describe merging and splitting of the Pomeron. With $y_{A, B}$ the rapidities of the external scattering particles, with $y_A \gg y_B$ and starting evolution at $y = y_B$,  $y_D$ is the rapidity of the splitting of the Pomeron into the two Pomeron state, $y_D \gg y_B$. Making use of \eqref{eq:N1} and evolving from $y_D$ to rapidities $y_B$ we have
\begin{align}
  \label{eq:evolv_to_YB}
   N_-^{(1)}(\Delta y_{DB}; \xt_1', \xt_2') & =- \frac{\asb}{2 \pi} 
\int \prod_{i=1}^2  d^2\yt'_i 
\int d^2 \yt_3' \frac{(\yt_1' - \yt_2')^2}{(\yt_1'- \yt_3')^2 (\yt_2'  - \yt_3')^2}
\notag \\
&
\tilde{g}(\Delta y_{DB}; \xt_1', \xt_2'; \yt_1', \yt_2')   \hat{N}_- (\yt_1', \yt_3')  \hat{N}_- (\yt_3', \yt_2'),
\end{align}
where we evaluated the factor quadratic in $ \hat{N}^{(0)}_-$ at rapidity $y_D$, i.e. the evolution starts at the triple Pomeron vertex and $\Delta y_{ij} = y_i - y_j$. Alternatively, we can take the two Pomeron state as our starting point and   evolve the system from $y_D$ towards $y_C$ (merging of the two Pomerons into a single Pomeron) and subsequent evolution up to $y_A$. We obtain for this contribution
\begin{align}
  \label{eq:evolv_to_YA}
  N_+^{(1)}(y_a - y_d; \xt_1, \xt_2) & =- \frac{\asb}{2 \pi} \int^{y_A}_{y_D} dy _C  
\int \prod_{i=1}^2  d^2\yt_i
  \tilde{g}(y_A-y_C; \xt_1, \xt_2; \yt_1, \yt_2)  \notag \\
& \hspace{-2cm}
\bigg[ \int d^2 \yt_3 \frac{(\yt_1 - \yt_2)^2}{(\yt_1 - \yt_3)^2 (\yt_2  - \yt_3)^2}
 \hat{N}^{(0)}_+ (y_C-y_D;\yt_1, \yt_3)  \hat{N}^{(0)}_+ (y_C-Y_D;\yt_3, \yt_2).
 \end{align}
When combining both expressions, we finally need to integrate over $y_D$  -- the rapidity of the lower triple Pomeron vertex. 
We therefore obtain the  Pomeron loop configuration through
\begin{align}
  \label{eq:M2}
  \mathcal{M}_2(Y; \xt_1, \xt_2; \xt_1', \xt_2') & = -i \int^{y_A}_{y_B} d y_D \left \langle  N_+^{(1)}(y_A - y_D, \xt_1, \xt_2) N_-^{(1)}(\Delta y_{DB}, \xt_1', \xt_2')    \right  \rangle.
\end{align}
To evaluate the  various integrals over rapidities  we use
\begin{align}
  \label{eq:rap_trick}
  \int_{y_B}^{y_A} dy_D \int _{y_D}^{y_A} dy_c = \int\limits_0^\infty d\Delta y_{AC} \int\limits_0^\infty d \Delta y_{CD} \int\limits_0^\infty d\Delta y_{DB} \delta(\Delta y_{AC} + \Delta y_{CD} + \Delta y_{DB} - Y) .
\end{align}
 We finally need the correlator of four color dipole amplitudes,  
\begin{align}
  \label{eq:8ptcorrelator}
 &   \int d^2 \yt_{3} \frac{(\yt_1 - \yt_2)^2}{(\yt_1 - \yt_3)^2 (\yt_2  - \yt_3)^2}
 \int d^2 \yt' \frac{(\yt_1' - \yt_2')^2}{(\yt_1' - \yt'_3)^2 (\yt_2'  - \yt'_3)^2} \notag \\
& \hspace{6cm} \bigg \langle  \hat{N}_+(\yt_1, \zt) \hat{N}_+ (\yt_3, \yt_2)    N_- (\yt_1 ', \yt'_3)  N_- (\yt'_3, \yt_2' ) 
\bigg \rangle \notag \\
& = 
 \frac{1}{2} \int d^2 \yt_{3} \frac{(\yt_1 - \yt_2)^2}{(\yt_1 - \yt_3)^2 (\yt_2  - \yt_3)^2} 
\int d^2 \yt' \frac{(\yt_1' - \yt_2')^2}{(\yt_1' - \yt'_3)^2 (\yt_2'  - \yt'_3)^2} \notag \\
& \hspace{3cm}
\left\langle  N_+ (\yt_1 , \yt_3) N_- (\yt_1 ', \yt'_3)  \right \rangle 
 \left
\langle  N_+ (\yt_2 , \yt_3) N_- (\yt_2 ', \yt'_3) \right \rangle,
 \end{align}
where we employed that  within the large $N_c$ approximation we only deal  with pairwise correlators of $\langle N_+ N_-\rangle$, see also \cite{Bartels:2009zc}. Since to order $g^2$ the dipole amplitude $\hat{N}(\xt_1, \xt_2)$ is symmetric under $\xt_1 \leftrightarrow \xt_2$, we have symmetry of the triple Pomeron vertex under switching the two dipoles  and we require corresponding symmetry factors. After contractions we therefore arrive at a symmetry factor $1/2$, in accordance with the discussion in \cite{Bartels:2007dm}. 
Making further use of the relation \cite{Braun:1997nu, Lipatov:1985uk}
\begin{align}
  \label{eq:derv_onE}
  \partial^2_{\yt_1} \partial^2_{\yt_2}E^{n,\nu}(\yt_{10},\yt_{20})& = \frac{16}{\yt_{12}^4} \left[\nu^2 + \frac{(n+1)^2}{4} \right]\left[\nu^2 + \frac{(n-1)^2}{4} \right]E^{n,\nu}(\yt_{10},\yt_{20}),
\end{align}
we arrive at
\begin{align}
  \label{eq:A2e}
   \mathcal{M}_2&(Y; \xt_1, \xt_2; \xt_1', \xt_2') =  \frac{-i\asb^2}{8 \pi^2 (2 \pi)^8}  \left( \frac{\asb^2}{4 N_c^2} \right)^2 \prod_{i=0}^3\int d\nu_i   \sum_{n_i=\text{even}} \int \frac{d\omega}{2 \pi i }
\frac{ e^{\omega Y}   }{\omega - \omega(n_0, \nu_0)} 
\notag \\
 & 
\frac{1}{\omega - \omega(n_1, \nu_1)-  \omega(n_2, \nu_2)} 
\frac{1}{\omega - \omega(n_3, \nu_3)}
\frac{(16 \nu_0^2 + 4 n_0^2)\left( \nu_1^2 + \frac{n_1^2}{4} \right)}{\left(\nu_1^2 + \frac{(n_1+1)^2}{4} \right) \left(\nu_1^2 + \frac{(n_1-1)^2}{4} \right)}
\notag \\
&
 \frac{(16 \nu^{2}_3 + 4 n_3^{2 })\left(\nu_2^2 + \frac{n_2^2}{4} \right)}{\left(\nu_2^2 + \frac{(n_2+1)^2}{4} \right) \left(\nu_2^2 + \frac{(n_2-1)^2}{4} \right)} 
 \prod_{l=0}^3 \int d^2\zt_l \prod_{i=1}^3 \int d^2 \yt_i \prod_{j=1}^3 \int d^2 \yt_j'   \notag \\
& 
 \frac{E^{n_0, \nu_0*}(\yt_1-\zt_0, \yt_2-\zt_0)  E^{n_1, \nu_1}(\yt_1-\zt_1, \yt_{3} - \zt_1)  E^{n_2, \nu_2}(\yt_{2} - \zt_2, \yt_{3}-\zt_2)}{\yt_{12}^2 \yt_{13}^2 \yt_{23}^2} \notag \\
& 
  \frac{E^{n_1, \nu_1*}(\yt_{1'}-\zt_1, \yt_{3'}-\zt_1)  E^{n_2, \nu_2* }(\yt_{2'}-\zt_2, \yt_{3'}-\zt_2) E^{n_3\nu_3}(\yt_{1'}-\zt_3, \yt_{2' }-\zt_3)}{\yt_{1'2'}^2 \yt_{1'3'}^2 \yt_{2'3'}^2} \notag \\
&
E^{n_0, \nu_0}(\xt_1 -\zt_0,\xt_2-\zt_0) E^{n_3, \nu_3*}(\xt_1' -\zt_3,\xt_2'-\zt_3).
\end{align}

\subsection{Determination of integrals}
\label{sec:integrals}

To evaluate the integrals over $d^2\yt_i, d^2\yt_i'$, $i=1, \ldots, 3$ we make use of a result by Korchemsky  \cite{Korchemsky:1997fy}. One has
\begin{align}
  \label{eq:V0_Korchemsky}
  V_0(\alpha, \beta, \gamma) & = \int \frac{d^2 \yt_1 d^2 \yt_2 d^2 \yt_3}{ \yt_{12}^2 \yt_{13}^2 \yt_{23}^2} E_{h_\alpha \bar{h}_{\alpha}}(\yt_{1\alpha},\yt_{2\alpha}) E_{h_\beta \bar{h}_{\beta}}(\yt_{2\beta},\yt_{3\beta}) E_{h_\gamma \bar{h}_{\gamma}}(\yt_{3\gamma},\yt_{1\gamma}) \notag \\
& =  \frac{\Omega(h_\alpha, h_\beta, h_\gamma) }{
y_{\alpha\beta}^{\Delta_{\alpha\beta} }
y_{\alpha\gamma}^{\Delta_{\alpha\gamma} }
y_{\beta\gamma}^{\Delta_{\beta \gamma} }
\bar{y}_{\alpha\beta}^{\bar{\Delta}_{\alpha\beta} }
\bar{y}_{\alpha\gamma}^{\bar{\Delta}_{\alpha\gamma} }
\bar{y}_{\beta\gamma}^{\bar{\Delta}_{\beta \gamma} }
} \notag \\
\Delta_{ij} & = h_i + h_j - h_k, \quad \bar{\Delta}_{ij}  = \bar{h}_i + \bar{h}_j - \bar{h}_k,  \quad k \neq i,j\,,
\end{align}
where $\bar{h}_i = 1-h_i^*$ and
\begin{align}
  \label{eq:E_mess}
   E_{h, \bar{h}}(\rt_1-\rt_0, \rt_2-\rt_0) & = 
 \left(\frac{r_{12}}{r_{10} r_{20}} \right)^{\frac{1+n}{2} + i \nu} 
 \left(\frac{r_{12}^*}{r_{10}^*  r_{20}^* } \right)^{\frac{1-n}{2} + i \nu}
=
 \left(\frac{r_{12}}{r_{10} r_{20}} \right)^{h} 
 \left(\frac{r_{12}^*}{r_{10}^*  r_{20}^* } \right)^{\bar{h}}, 
\notag \\
  E_{h, \bar{h}}^*(\rt_1-\rt_0, \rt_2-\rt_0) & =  E_{\bar{h}^*, {h}^*}(\rt_1-\rt_0, \rt_2-\rt_0).
\end{align}
%%% check this i.e. in comparison to Eq.~\eqref{eq:E} one has $n \to -n$. Since we will set $n=0$ for our current calculation, this is not of direct concern, but we annotate it for future investigations. 
We therefore obtain
\begin{align}
  \label{eq:V456}
  V_0(0;1,2) & = 
 \frac{\Omega(\bar{h}_0^*, h_1, h_2) e^{i\phi_{01}(n_1 - n_2 - n_0)} e^{i\phi_{02}(n_2 - n_1 - n_0)} e^{i\phi_{12}(n_1 + n_2 +n_0)}}{(\zt_{01}^2)^{\frac{1}{2} + i(\nu_1 -\nu_2 - \nu_0)} (\zt_{02}^2)^{\frac{1}{2} + i (\nu_2 - \nu_1 -  \nu_0)} (\yt_{12}^2)^{\frac{1}{2} + i (\nu_0+\nu_2 + \nu_1)}}, \notag \\
& 
\notag \\
V_0(1,2;3) & = 
 \frac{\Omega(\bar{h}_1^*, \bar{h}_2^*, h_3) e^{-i\phi_{12}(n_1 + n_2 + n_3)} e^{i\phi_{13}(n_3 - n_1 +n_2)} e^{i\phi_{23}(n_3 - n_2 +n_1)}}{(\zt_{12}^2)^{\frac{1}{2} - i(\nu_1 +\nu_2 + \nu_3)} (\zt_{13}^2)^{\frac{1}{2} + i (\nu_3 - \nu_1 +  \nu_2)} (\zt_{23}^2)^{\frac{1}{2} + i (\nu_3-\nu_2 + \nu_1)}} ,
\end{align}
where $z_{ij} = |z_{ij}| e^{i\phi_{ij}}$ and
\begin{align}
  \label{eq:hlist}
  h_0& = \frac{1+n_0}{2}+i\nu_0, &  
  h_1& = \frac{1+n_1}{2}+i\nu_1,  \notag \\
  h_2& = \frac{1+n_2}{2}+i\nu_2,  &
   h_3& = \frac{1+n_3}{2}+i\nu_3 .
\end{align}
We  then arrive at 
\begin{align}
  \label{eq:A2f}
   \mathcal{M}_2&(Y; \xt_1, \xt_2; \xt_1', \xt_2')   = -i  \frac{\asb^6}{128 \pi^{10} N_c^4}        \int\limits_{\delta - i \infty}^{\delta + i \infty}  \frac{d\omega}{2 \pi i }   \prod_{i=0}^3 \int\limits_{-\infty}^\infty  d\nu_i \sum_{n_i = \text{even}}  
\frac{e^{\omega Y} ( \nu_0^2 +  \frac{n_0^2}{4})}{\omega - \omega(n_0, \nu_0)}  
 \notag \\
 & 
\frac{( \nu_3^{2} +  \frac{n_3^{2 }}{4})}{\omega - \omega(n_3, \nu_3)} \frac{ \Omega(\bar{h}_0^*, h_1, h_2) \Omega(\bar{h}_1^*, \bar{h}_2^*, h_3)}{\omega - \omega(n_1, \nu_1)-  \omega(n_2, \nu_2)} 
\prod_{j=1}^2 \frac{\left( \nu_j^2 + \frac{n_j^2}{4} \right)}{\left(\nu_j^2 + \frac{(n_j+1)^2}{4} \right) \left(\nu_j^2 + \frac{(n_j-1)^2}{4} \right)}  \notag \\
& 
\prod_{i=0}^3 \int d^2 \zt_i 
\frac{E^{n_0, \nu_0}(\xt_1-\zt_0,\xt_2-\zt_0) E^{n_3, \nu_3*}(\xt_1' -\zt_3,\xt_2'-\zt_3)}{(\zt_{01}^2)^{\frac{1}{2} + i(\nu_1 -\nu_2 - \nu_0)} (\zt_{02}^2)^{\frac{1}{2} + i (\nu_2 - \nu_1 -  \nu_0)}}
 \notag \\
&  \qquad 
\frac{  e^{i\phi_{01}(n_1 - n_2 - n_0)} e^{i\phi_{02}(n_2 - n_1 - n_0)} e^{i\phi_{12}(n_0-n_3)}  e^{i\phi_{13}(n_3 - n_1 +n_2)} e^{i\phi_{23}(n_3 - n_2 +n_1)}}{ (\zt_{12}^2)^{1 + i (\nu_0- n_3)}  (\zt_{13}^2)^{\frac{1}{2} + i (\nu_3 - \nu_1 +  \nu_2)} (\zt_{23}^2)^{\frac{1}{2} + i (\nu_3-\nu_2 + \nu_1)} } .
\end{align}
In the following we are mainly interested in the case  $n=n'=n_1 = n_2 =0$, which is the dominant configuration in the evaluation of the BFKL Green's function Eq.~\eqref{eq:f_Lev}; $\omega(n, \nu)<0$ for $n \neq 0$ and contributions with finite $n$ are therefore increasingly suppressed in the high energy limit. The study of contributions with finite conformal spin is then left as a task for future research. In this case, the evaluation of the remaining coordinate integrals is  relatively straightforward. Since the resulting integral is divergent, we introduce dimensional regularization in $d= 2 + 2 \epsilon$ dimensions\footnote{Even though original expression was not obtained in $2 + 2 \epsilon$ dimensions, this is justified since the final result will be free of divergences} and with  $\zt_i \to \zt_i + \zt_0$ we have 
 \begin{align}
    \label{eq:J00g}
    J_{00} & = \int
\frac{ d^{2+ 2 \epsilon} \zt_1 d^{2 + 2 \epsilon} \zt_2}{(\zt_{01}^2)^{1 + h_1 - h_2 - h_0} 
(\zt_{02}^2)^{1 + h_2 - h_2 - h_0}  }
\frac{1}{(\zt_{12}^2)^{1 + h_0 - h_3} (\zt_{13}^2)^{h_3-h_1 + h_2} (\zt_{23}^2)^{h_3- h_2 + h_1}
}.
  \end{align}
 Using
 \begin{align}
    \label{eq:J00g2}
&
\int d^{2 + 2 \epsilon} \zt_1
\frac{1}{(\zt_{01}^2)^{1 + h_1 - h_2 - h_3} 
(\zt_{12}^2)^{1 + h_0 - h_3} (\zt_{13}^2)^{h_3-h_1 + h_2} 
}
\notag \\
   & = \pi ^{\epsilon +1}   \left(\zt_{23}^2\right){}^{-h_0 +h _1-h _2-\epsilon }   \left(\zt_{02}^2\right){}^{h_3 -h _1+h_2+\epsilon -1}
   \left(\zt_{03}^2\right){}^{-h_3+h_0 +\epsilon } \notag \\
&  \qquad  \frac{ \Gamma \left(h_0 +\epsilon -h _1+h _2\right)
  \Gamma
   \left(-\epsilon +h _1-h _2-h_3+1\right) \Gamma \left(-h_0 -\epsilon
   +h_3\right)}{\Gamma \left(-h_0 +h
   _1-h _2+1\right) \Gamma \left(h_0 -h_3+1\right) \Gamma \left(-h
   _1+h _2+h_3\right)},
\end{align}
we have
\begin{align}
  \label{eq:J_oo}
  J_{00} & = \frac{\pi ^{\epsilon +2} \Gamma (1-\epsilon ) \Gamma \left(h_0 +\epsilon -h
   _1+h _2\right) \Gamma \left(-h_0 -h_3+1\right) \Gamma \left(-\epsilon
   +h _1-h _2-h_3+1\right)}{\Gamma \left(-h_0 +h
   _1-h _2+1\right) \Gamma (2 \epsilon ) \Gamma \left(h_0 -h_3+1\right)
   } \notag \\
& 
\frac{ \Gamma \left(-h_0 -\epsilon +h_3\right) \Gamma \left(h_0 +2 \epsilon +h_3-1\right)
   \left(\zt_{03}^2\right){}^{-h_3+h_0 +2 \epsilon -1}}{\Gamma \left(-h _1+h _2+h_3\right) \Gamma \left(-h_0 -\epsilon
   -h_3+2\right) \Gamma \left(h +\epsilon +h_3\right)}.
\end{align}
We find that $J_{00} \to 0$ for $\epsilon \to 0$ and  $\zt_{03}^2 $ finite, while the expression diverges  for $\zt_{03}^2 \to 0$. The above result is therefore proportional to a delta function. To evaluate its coefficient we integrate $J_{00}$  and find 
\begin{align}
  \label{eq:intJ00}
  \int d^{2+2\epsilon} \zt_3 J_{00} & = \delta(\nu_0 - \nu_3) \frac{ \pi^4}{2\nu_0^2}+ \mathcal{O}(\epsilon),
\end{align}
and therefore
\begin{align}
  \label{eq:J00result}
  J_{00} & = \delta^{(2)}(\zt_{03}) \delta(\nu_0 - \nu_3) \frac{\pi^4 }{2\nu_0^2}.
\end{align}
In the approximation $n_i = 0$, $i = 0, \ldots, 3$ we therefore find
\begin{align}
  \label{eq:A2f0}
   \mathcal{M}_2& (Y; \xt_1, \xt_2; \xt_1', \xt_2')  \simeq 
-i  \frac{\asb^6}{256 \pi^{6} N_c^4}        \int\limits_{\delta - i \infty}^{\delta + i \infty}  \frac{d\omega}{2 \pi i } e^{\omega Y}   \prod_{i=0}^2 \int\limits_{-\infty}^\infty  d\nu_i 
\frac{ 1}{[\omega - \omega(0, \nu_0)]^2} 
 \notag \\
 & 
\frac{ \Omega\left(\tfrac{1}{2} - i \nu_0 , \tfrac{1}{2} + i \nu_1, \tfrac{1}{2} + i \nu_2 \right)
\Omega\left(\tfrac{1}{2} + i \nu_0 , \tfrac{1}{2} - i \nu_1, \tfrac{1}{2} - i \nu_2 \right) }{\omega - \omega(n_1, \nu_1)-  \omega(n_2, \nu_2)} 
\frac{\nu_1^2 \nu_2^2 \left(\nu_0^2 + \tfrac{1}{4} \right)^2}{\left(\nu_1^2 + \tfrac{1}{4} \right)^2 \left(\nu_2^2 + \tfrac{1}{4} \right)^2}
\notag \\
& 
 \int d^2 \zt_0  \frac{\nu_0^2}{\left(\nu_0^2 + \frac{1}{4}\right)^2}
E^{0, \nu_0}(\xt_1-\zt_0,\xt_2-\zt_0) E^{0, \nu_0*}(\xt_1' -\zt_0,\xt_2'-\zt_0)
\notag \\
 =&
-i  \frac{\asb^6}{256 \pi^{6} N_c^2}     \prod_{i=0}^2 \int\limits_{-\infty}^\infty  d\nu_i 
\frac{e^{Y [\omega(\nu_1) + \omega(\nu_2)]} + e^{Y \omega(\nu_0)}  
\left[Y\left(\omega(\nu_0)  - \omega(\nu_1) -\omega(\nu_2) \right)-1 \right]}{ \left(\omega(\nu_0)  - \omega(\nu_1) - \omega(\nu_2) \right)^2}
 \notag \\
 & 
%\frac{
 \Omega\left(\tfrac{1}{2} - i \nu_0 , \tfrac{1}{2} + i \nu_1, \tfrac{1}{2} + i \nu_2 \right)
\Omega\left(\tfrac{1}{2} + i \nu_0 , \tfrac{1}{2} - i \nu_1, \tfrac{1}{2} - i \nu_2 \right) %}{\omega - \omega(n_1, \nu_1)-  \omega(n_2, \nu_2)} 
\frac{\nu_1^2 \nu_2^2 \left(\nu_0^2 + \tfrac{1}{4} \right)^2}{\left(\nu_1^2 + \tfrac{1}{4} \right)^2 \left(\nu_2^2 + \tfrac{1}{4} \right)^2}
\notag \\
& 
 \int d^2 \zt_0  \frac{\nu_0^2}{\left(\nu_0^2 + \frac{1}{4}\right)^2}
E^{0, \nu_0}(\xt_1-\zt_0,\xt_2-\zt_0) E^{0, \nu_0*}(\xt_1' -\zt_0,\xt_2'-\zt_0),
\end{align}
where we used $\omega(\nu) \equiv \omega(0,\nu)$. Now, this can be 
  compared to the single Pomeron contribution in the $n=0$ approximation,
\begin{align}
  \label{eq:iM1}
  \mathcal{M}_1 & =  i \frac{\bar{\alpha}_s^2}{4N_c^2}   \int\limits_{-\infty}^\infty d\nu \int d^2 \xt_0 
 \frac{\nu^2}{\left(\nu^2 + \frac{1}{4} \right) \left(\nu^2 + \frac{1}{4} \right)} 
%\notag \\
%& \hspace{2cm}
e ^{Y \omega(\nu)}
E^{0,\nu}(\xt_{10}, \xt_{20})E^{0,\nu,*}(\xt_{1'0}, \xt_{2'0}).
\end{align}
The result allows us to compare already with Eq.~(44) and Eq.~(47) of \cite{Braun:2013tha}. We find 3 important differences. First of all, the authors of \cite{Braun:2013tha} simplify integrals over $\nu$ through $\int_{-\infty}^\infty d \nu_i \to 2 \int_{0}^\infty d \nu $. To the best of our understanding, the vertex $\Omega$ does however not allow for such a simplification. Second, we find a different structure in the $Y$ dependent exponents, after evaluation of the $\omega$ integral i.e. to recover our result, the second term in the last line of Eq.~(47) of \cite{Braun:2013tha} should be a $e^{\omega(\nu)Y}$ and not $e^{\omega(\nu_1)Y}$; we believe that this is merely a typing error since the evaluation of the $\omega$ integral is relatively straightforward and we find agreement before the evaluation of the integral over $\omega$. Finally, we find a relative suppression of the Pomeron loop result  with respect to the result presented in \cite{Braun:2013tha} by a factor of $1/32$. We believe that this difference is due to the particular normalization chosen for the BFKL Green's function taken in \cite{Braun:2013tha} which, to the best of our understanding, is not in agreement with the result presented in \cite{Lipatov:1985uk}. It is then in particular this relative factor of $1/32$ which give rise to the suppression of the Pomeron loop with respect to the result of \cite{Braun:2013tha}. 

Finally, including the wave functions of external scattering particles we have for the single Pomeron exchange contribution,
 \begin{align}
  \label{eq:scattering_cont33}
   \mathcal{A}_1(s,t=0) & =  i \frac{s}{Q_A Q_B} \cdot  \left(\frac{\asb^2}{8} \right) 
\int d\nu e^{Y \omega( \nu)} H^2(\nu)  \left(\frac{Q_A^2}{Q_B^2} \right)^{i \nu},
\end{align}
while the  Pomeron-loop contribution reads
\begin{align}
  \label{eq:scattering_cont4}
   \mathcal{A}_2(s,t=0) & = - i \frac{s}{Q_A Q_B} \cdot  \left(\frac{\asb^2}{8} \right)  \frac{\asb^4}{64 N_c^2 \pi^6} 
\int d\nu  \int d\nu_1 \int  d\nu_2  \, \,H^2(\nu)  \left(\frac{Q_A^2}{Q_B^2} \right)^{i \nu} \notag \\
& \hspace{-1cm}\frac{e^{Y [\omega(\nu_1) + \omega(\nu_2)]} + e^{Y \omega(\nu)}  \left[Y\left(\omega(\nu)  - \omega(\nu_1) - \omega(\nu_2) \right)-1 \right]}{ \left(\omega(\nu)  - \omega(\nu_1) - \omega(\nu_2) \right)^2} \frac{\left(\nu^2 + \frac{1}{4} \right)^2 \nu_1^2 \nu_2^2}{\left(\nu_1^2 + \frac{1}{4} \right)^2 \left(\nu_2^2 + \frac{1}{4} \right)^2}
\notag \\
 & \hspace{-1cm} 
\Omega\left(\frac{1}{2}-i \nu, \frac{1}{2} + i \nu_1, \frac{1}{2} + i \nu_2\right)
\Omega\left(\frac{1}{2}-i \nu_1, \frac{1}{2}-i \nu_2, \frac{1}{2}+i \nu\right).
\end{align}
Through the optical theorem we have furthermore
\begin{align}
  \label{eq:sigma}
  \sigma_{\text{tot}, i}(\gamma^*_A\gamma^*_B \to X) & = \frac{1}{s} \Im\text{m} \mathcal{A}_i(s,0), & i & =1,2,
\end{align}
for the contribution of single Pomeron exchange and the Pomeron loop to the total cross-section, which we will study in the following.

\subsection{Explicit expressions for $\Omega$}
\label{sec:dipole_final}
To keep this paper self-contained, we further provide the explicit expressions for  $\Omega$. We have 
\begin{align}
    \label{eq:Omega}
    \Omega(h_\alpha, h_\beta, h_\gamma;\bar{h}_\alpha, \bar{h}_\beta, \bar{h}_\gamma) & = \frac{\pi^3 \sum_{i=1}^3 J_i(h_\alpha, h_\beta, h_\gamma) 
    \bar{J}_i(\bar{h}_\alpha, \bar{h}_\beta, \bar{h}_\gamma)}{\Gamma^2(h_\alpha) \Gamma^2(h_\beta)\Gamma(1-h_\alpha)\Gamma(1-h_\beta)\Gamma(1-h_\gamma)} .
\end{align}
The functions $J_i, \bar{J}_i$, $i=1, \ldots 3$ have been obtained in \cite{Korchemsky:1997fy} and read:
\begin{align}
    \label{eq:J1}
   J_1(h_\alpha, h_\beta, h_\gamma)  
   &=
   \Gamma \left(1-h_{\alpha }\right) \Gamma \left(h_{\alpha }\right) \Gamma
   \left(1-h_{\beta }\right) \Gamma \left(h_{\beta }\right) \Gamma \left(h_{\alpha }+h_{\beta }-h_{\gamma }\right) \notag \\
   & \qquad \cdot
   \int_0^1 dx \, 
   (1-x)^{-h_{\gamma }} \,
   _2F_1\left(1-h_{\alpha },h_{\alpha };1;x\right) \, _2F_1\left(1-h_{\beta },h_{\beta
   };1;x\right) \notag \\
   & \hspace{-1cm}=
   \frac{\Gamma \left(1-h_{\alpha }\right) \Gamma \left(h_{\alpha }\right) \Gamma
   \left(h_{\alpha }+h_{\beta }-h_{\gamma }\right) }{\Gamma \left(1-h_{\beta }\right) \Gamma \left(h_{\beta }\right)} G_{4,4}^{2,4}\left(1\left|
\begin{array}{c}
 1-h_{\beta },h_{\beta },h_{\gamma },h_{\gamma } \\
 0,0,h_{\gamma }-h_{\alpha },h_{\alpha }+h_{\gamma }-1 \\
\end{array}
\right.\right),
\end{align}
where the representation in terms of a Meijer G function can be obtained through employing  the Mellin-Barnes representation of the hypergeometric function and subsequent use of the second Barnes lemma, see e.g. \cite{Smirnov:2006ry}. One further has
\begin{align}
 \label{eq:J2}
 J_2(h_\alpha, h_\beta, h_\gamma)  &=\frac{\Gamma \left(1-h_{\alpha }\right) \Gamma \left(h_{\alpha }\right) \Gamma
   \left(1-h_{\beta }\right) \Gamma \left(h_{\beta }\right) \Gamma \left(1-h_{\gamma
   }\right){}^2 \Gamma \left(h_{\alpha }+h_{\beta }-h_{\gamma }\right) \,
   }{\Gamma \left(-h_{\alpha
   }-h_{\gamma }+2\right) \Gamma \left(h_{\alpha }-h_{\gamma }+1\right)} \notag \\
   & \qquad  _4F_3\left(1-h_{\beta },h_{\beta },1-h_{\gamma },1-h_{\gamma };1,-h_{\alpha
   }-h_{\gamma }+2,h_{\alpha }-h_{\gamma }+1;1\right), \notag \\
   J_3(h_\alpha, h_\beta, h_\gamma) & = J_2(h_\beta, h_\alpha, h_\gamma).
\end{align}
As well as 
\begin{align}
    \label{eq:J1b}
 \bar{J}_1(\bar{h}_\alpha, \bar{h}_\beta, \bar{h}_\gamma) & =  (-)^{n_\alpha + n_\beta} \frac{\Gamma \left(1-\bar{h}_{\alpha }\right){}^2 \Gamma \left(\bar{h}_{\alpha }\right)
   \Gamma \left(1-\bar{h}_{\beta }\right){}^2 \Gamma \left(\bar{h}_{\beta }\right)
  }{\Gamma
   \left(1-\bar{h}_{\gamma }\right) \Gamma \left(-\bar{h}_{\alpha }-\bar{h}_{\beta
   }+\bar{h}_{\gamma }+1\right)}
   \notag \\
   & \hspace{-2cm}
   \int_0^1 dx \,  x^{-\bar{h}_{\gamma }}  (1-x)^{-\bar{h}_{\alpha }-\bar{h}_{\beta }+\bar{h}_{\gamma }}\, _2F_1\left(1-\bar{h}_{\alpha },1-\bar{h}_{\alpha
   };1;x\right) \, _2F_1\left(1-\bar{h}_{\beta },1-\bar{h}_{\beta };1;x\right).
\end{align}
Note that the overall factor $(-)^{n_\alpha + n_\beta}$ violates the separability into functions which depend on $h_i$ ($\bar{h}_i$), $i=1, \ldots 3$ only. Since $n_i = h_i - \bar{h}_i$ this separability can be however easily re-obtained, if desired. As already pointed out in \cite{Braun:2013tha}, the integral Eq.~\eqref{eq:J1b} cannot be expressed in terms of a Meijer G function, as provided in \cite{Korchemsky:1997fy}. It is however possible to express the function through a Mellin-Barnes representation,
\begin{align}
    \label{eq:J1bMB}
 \bar{J}_1(\bar{h}_\alpha, \bar{h}_\beta, \bar{h}_\gamma) 
 & =
 (-)^{n_\alpha + n_\beta}
\int \frac{ds}{2 \pi i} \int \frac{dt}{2 \pi i} 
\frac{\Gamma (-s) \Gamma (-t) \Gamma \left(s-\bar{h}_{\alpha }+1\right){}^2 \Gamma
   \left(-s+2 \bar{h}_{\alpha }-1\right) }{\Gamma \left(\bar{h}_{\alpha }\right)
   \Gamma \left(\bar{h}_{\beta }\right) }
   \notag \\
   & \qquad
   \frac{\Gamma \left(t-\bar{h}_{\beta }+1\right){}^2
   \Gamma \left(-t+2 \bar{h}_{\beta }-1\right) \Gamma \left(s+t-\bar{h}_{\alpha
   }-\bar{h}_{\beta }+\bar{h}_{\gamma }+1\right)}{\Gamma \left(-\bar{h}_{\alpha }-\bar{h}_{\beta
   }+\bar{h}_{\gamma }+1\right) \Gamma \left(s+t-\bar{h}_{\alpha }-\bar{h}_{\beta}+2\right)},
 \end{align}
 where the contours run to the right of all left running poles and to the left of all right running poles. Finally one has
 \begin{align}
     \label{eq:J2b}
   \bar{J}_2(\bar{h}_\alpha, \bar{h}_\beta, \bar{h}_\gamma)    &= 
    (-)^{n_\alpha} \frac{\Gamma \left(1-\bar{h}_{\alpha }\right) \Gamma \left(1-\bar{h}_{\beta }\right){}^2
   \Gamma \left(\bar{h}_{\beta }\right) \Gamma \left(\bar{h}_{\gamma }\right) \,
  }{\Gamma \left(-\bar{h}_{\alpha }-\bar{h}_{\beta
   }+\bar{h}_{\gamma }+1\right)}
   \notag \\
   & \hspace{-2cm}
   \int_0^1 dx \,  _2F_1\left(1-\bar{h}_{\alpha },1-\bar{h}_{\alpha };1;x\right) \,
   _2F_1\left(1-\bar{h}_{\beta },1-\bar{h}_{\beta };1;x\right) (1-x)^{-\bar{h}_{\alpha
   }-\bar{h}_{\beta }+\bar{h}_{\gamma }}\notag \\
   &  \hspace{-2cm}  =(-)^{n_\alpha} 
   \frac{\Gamma \left(1-\bar{h}_{\alpha }\right) \Gamma \left(\bar{h}_{\gamma }\right)
  }{\Gamma \left(\bar{h}_{\beta }\right) \Gamma \left(-\bar{h}_{\alpha
   }-\bar{h}_{\beta }+\bar{h}_{\gamma }+1\right)}  G_{4,4}^{4,2}\left(1\left|
\begin{array}{c}
 1-\bar{h}_{\beta },\bar{h}_{\beta },\bar{h}_{\gamma },\bar{h}_{\gamma } \\
 0,0,\bar{h}_{\gamma }-\bar{h}_{\alpha },\bar{h}_{\alpha }+\bar{h}_{\gamma }-1 \\
\end{array}
\right.\right), \notag \\
\bar{J}_3(\bar{h}_\alpha, \bar{h}_\beta, \bar{h}_\gamma)  & = \bar{J}_2(\bar{h}_\beta, \bar{h}_\alpha, \bar{h}_\gamma) .
 \end{align}
To evaluate the above functions it is either possible to evaluate numerically the integrals over the products of two hypergeometric functions or to make use of the representation in terms of the Meijer G functions and/or the double Mellin-Barnes integral. While the former is faster, we find problems in the numerical precision of our codes, as soon as one reaches values $|\nu_0|, |\nu_1|, |\nu_2| > 1$. For the current study, the function $\Omega = \Omega(\nu, \nu_1, \nu_2)$ has been determined first for a grid $[-3.5, 3.5] \times [-3.5, 3.5] \times [-3.5, 3.5]$ through \texttt{Wolfram Mathematica}, using the representation in terms of the Meijer G functions and/or the double Mellin-Barnes integral, see also \cite{Karina} for different attempts in evaluating the Pomeron loop correction numerically and a detailed discussion of the advantages of different approaches.

Numerical convergence was further  assessed at both the level of the triple-Pomeron vertex and of the final multidimensional integrals using a \texttt{Fortran} code. As a first pointwise test of the Mellin--Barnes implementation, the vertex was evaluated along the diagonal configuration $\nu,  \nu_1, \nu_2$ over an extended range. The calculation reproduces the known value $\Omega(0,0,0)=7766.6790118$ and yields a smooth and rapidly decreasing tail, with $|\Omega(t,t,t)|^2=9.50\times 10^2$, $1.56\times 10^1$, $1.23$, $1.84\times 10^{-1}$, $3.79\times 10^{-2}$, $9.14\times 10^{-3}$, $2.32\times 10^{-3}$, and $5.52\times 10^{-4}$ for $ t=1, \ldots, 8$, respectively. The Mellin--Barnes convergence criteria were satisfied through $t=8$. For the full three-dimensional integration, convergence was then tested by varying the cutoff from $\nu_{max}$=3.5 to 4.0 and 5.0. Increasing the cutoff from 3.5 to 4.0 changes %I$^2$​, and consequently
$-\sigma_2/\sigma_1$, by at most $3.0\times 10^{-5}$ over 4$\le$ Y$\le$ 16, while extending the integration domain further to $\nu_{max}$ produces a smaller maximum variation of $2.3 \cdot 10^{-5}$. The single-Pomeron contribution is even less sensitive to the cutoff, with relative variations below $7\times 10^{-8}$. We consequently adopt $\nu_{max}=4$ as the  cutoff, for which all independent Mellin--Barnes vertices satisfy the prescribed convergence criteria, while the extended pointwise test confirms the stability and decay of the vertex itself up to $|\nu|=8$.

\section{Numerical results}
\label{sec:pheno}

\begin{figure}[th]
  \centering
\includegraphics[width=\textwidth]{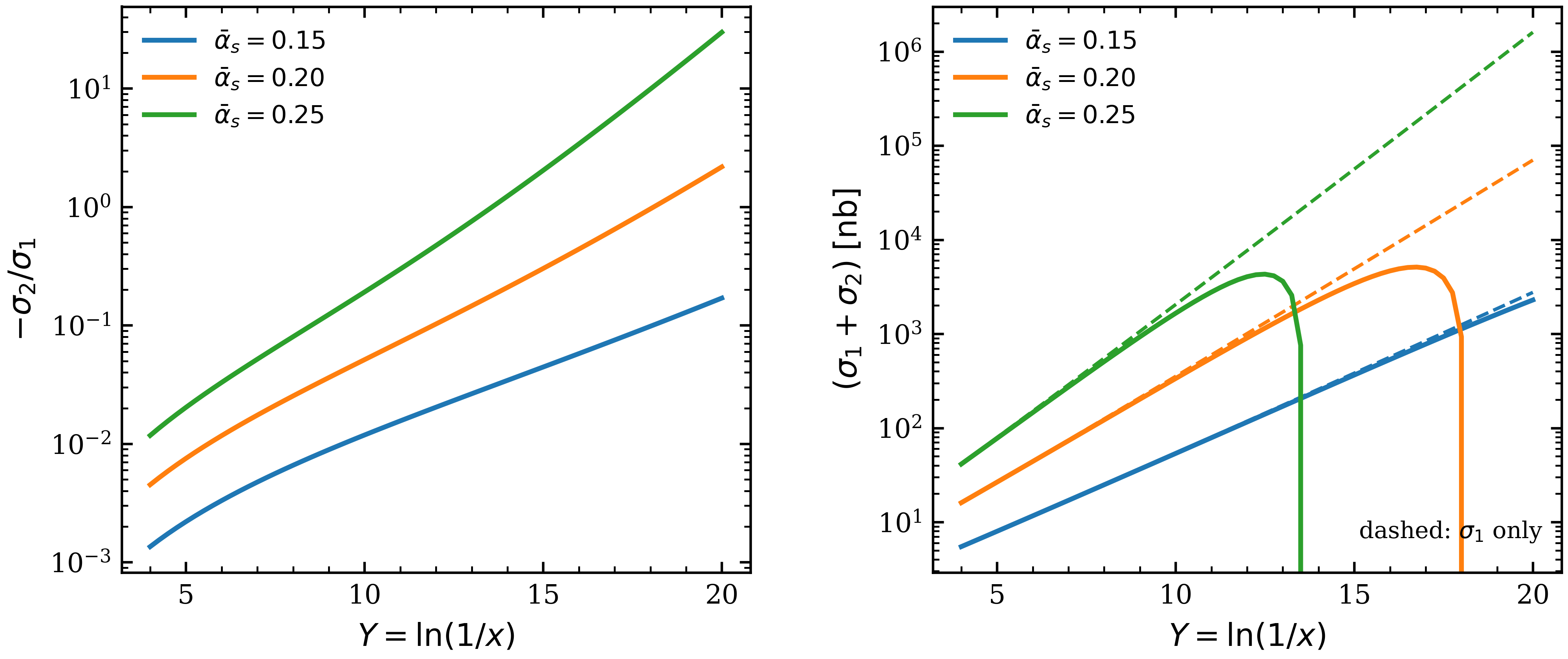}
  \caption{Ratio of the Pomeron-loop configuration over single Pomeron exchange vs $Y$ (left) and single Pomeron exchange combined with the Pomeron loop correction (right); dashed lines show single Pomeron exchange only.}
  \label{fig:vanilla}
\end{figure}  
In the following we finally evaluate the numerical size of the
Pomeron-loop configuration with respect to single Pomeron
exchange. While phenomenologically relevant cases are either Deep
Inelastic Scattering of a virtual photon on a proton or exclusive
photoproduction of heavy vector mesons, they require modeling and/or
fits of the hadronic impact factor which provides an additional source
of uncertainty. We therefore leave such explorations for a future
study, and consider here scattering of two virtual photons with virtualities $Q_{A}^2, Q_B^2$ at high
center-of-mass energy $\sqrt{s}$.  The dependence on energy is  encoded through through $Y  = \ln 1/x$, $x=Q_A^2/s$, which allows to relate the evolution parameter $Y$ to parton momentum fractions probed in experiment.

For the
first set of studies we consider a hierarchy of virtualities,
i.e. $Q^2_A = 25$~GeV$^2$, while $Q_B^2 = 2$~GeV$^2$, with
$L = \ln Q_A^2/Q_B^2 \simeq 2.53$. The strong coupling constant is 
evaluated at the scale of the harder photon $\asb(Q_A^2) =
0.2$, if  not indicated otherwise. Dependence on the value of the strong coupling constant is then investigated at first in Fig.~\ref{fig:vanilla}, left which shows the ratio of the Pomeron loop
correction ($\sigma_2$) over the single Pomeron exchange ($\sigma_1$).

Since the ratio is directly proportional to the fourth power of the strong coupling constant, it is no surprise that the relative size of the Pomeron loop contribution depends strongly on the value of $\alpha_s$. Apart from the overall normalization,  the growth of the ratio with $Y = \ln 1/x$ is also more pronounced for larger values of the strong coupling constant.  While the
Pomeron loop contribution is in general small for small to moderate
values of $Y$, the ratio exceeds one for very large values of $Y$. For our studies, this is the case for values  of $Y = 15-20$,  which correspond to center of mass energies which cannot be reached by today's collider experiments. However, such energies are potentially accessible in cosmic rays, as well as at the planned “Future Circular Collider” in the hadron-hadron mode (FCC-hh) \cite{FCC:2018vvp}, where one aims at  center of mass energies of  $\sqrt{s} = 100$~TeV.

For currently accessible collider energies we find for 
$x=10^{-6}$, corresponding to $Y \simeq 13.8$, a loop correction which
amounts to approximately $20\%$ of the single Pomeron exchange cross-section, while for  $x=10^{-5}$, corresponding to
$Y \simeq 11.5$, the correction reduces to $9\%$. We stress again that these values depend
strongly on the value taken for the overall factor of
$\asb^4$. The right panel of Fig.~\ref{fig:vanilla} shows the combination of
single Pomeron exchange and Pomeron-loop configuration, $\sigma_1 + \sigma_2$. While the growth of the cross-section is dominated by single Pomeron exchange, the effects of the Pomeron loop change this behavior dramatically, once the Pomeron loop becomes  similar in size than the single Pomeron exchange.

\begin{figure}[th]
  \centering
  \includegraphics[width = .47\textwidth]{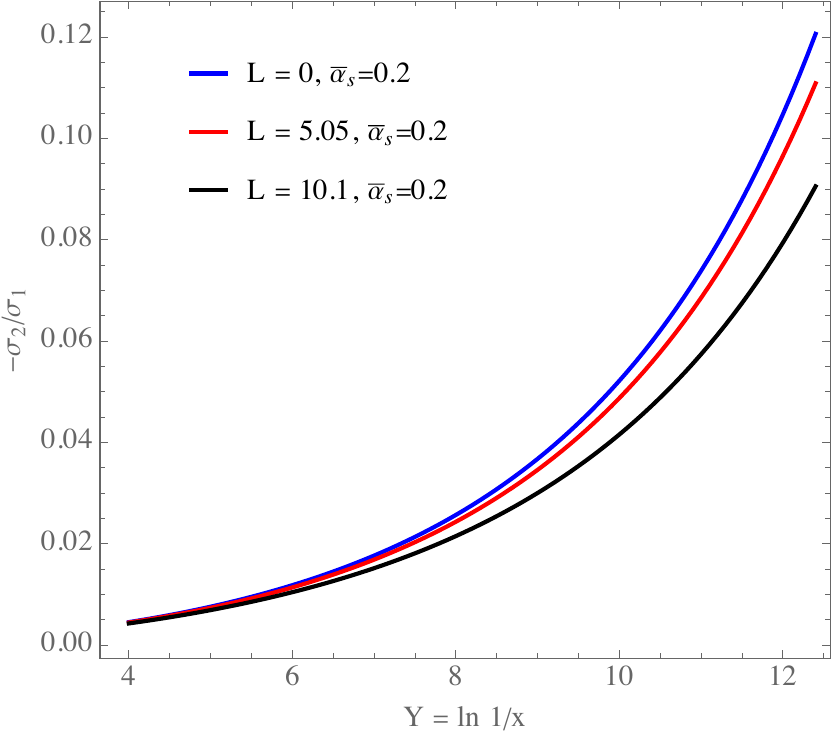}
  \includegraphics[width = .47\textwidth]{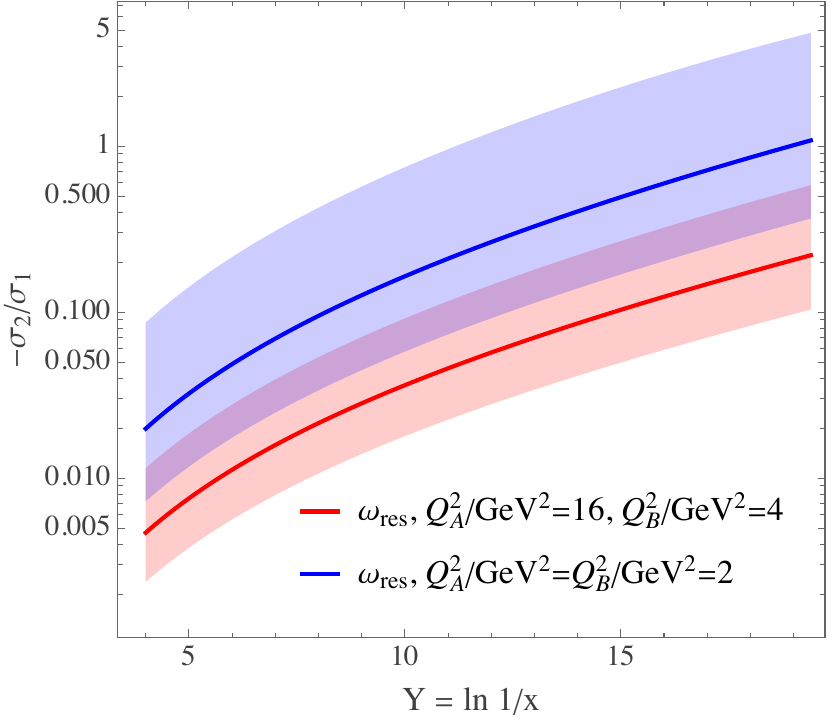}
  
  \caption{Left: Ratio of the Pomeron-loop configuration over single Pomeron exchange vs $Y$ for different values of $L = \ln Q_A^2/Q_B^2$. Right: Ratio of the Pomeron-loop configuration over single Pomeron exchange vs $Y$ for a resummed BFKL eigenvalue.}
  \label{fig:vanilla_l1}
\end{figure}

The left panel of  Fig.~\ref{fig:vanilla_l1} finally considers the effect of variations in the parameter $L = \ln Q_A^2/Q_B^2$, while we keep the value of the running coupling fixed. Even though one might expect a strong suppression of the Pomeron loop for a strong hierarchy $Q_A \gg Q_B$, our result does not provide such a behavior. While the cross-sections itself strongly depend on the chosen values of the photon virtualities, one finds that the ratio of Pomeron loop configuration over single Pomeron exchange depends only weakly on variations of $L$, if the value of the running coupling constant is kept constant.

As it is well known, the leading order BFKL eigenvalue,
\begin{align}
  \label{eq:omegaEVLO}
  \omega(\nu) & = \asb \chi_0\left(\frac{1}{2} + i \nu, 0 \right),
\end{align}
gives in general  rise to a growth with energy of the single Pomeron exchange which is too strong; for single Pomeron exchange one finds $\sigma_1 \sim e^{\lambda y}$ where the parameter $\lambda$ can be fitted to $\lambda \simeq 0.532$ if $\asb=0.2$.  A rate of growth which is in accordance with e.g. DIS data can  however be achieved if one takes into account the complete NLO corrections, together with a resummation of collinear logarithms and  an appropriate treatment of large $\beta_0$ terms in the NLO BFKL eigenvalue, see e.g. \cite{Hentschinski:2012kr, Hentschinski:2013id}.   Since a strong rise of the single Pomeron contribution will naturally further  enhance the magnitude of the Pomeron-loop configuration, taking into account such effects is of interest for the following exploration. Indeed \cite{Dumitru:2007ew} concluded within the framework of a 1+1 dimensional model that running coupling effects lead to  a strong suppression of Pomeron loop effects. 
While including  the complete NLO corrections is beyond the scope of this paper, an estimate can be made through a BLM resummed strong coupling constant and through imposing the so-called kinematic constraint onto the leading order BFKL eigenvalue. Following \cite{Hentschinski:2012kr, Hentschinski:2013id} we use to this end
\begin{align}
  \label{eq:omega_resummed}
  \omega_{res.}(\nu) & = \asbt(\gamma)  \chi_0\left(\gamma, 0 \right) \notag \\
  &
  \hspace{.5cm} + \sum_m \left[ \frac{1-\gamma + m}{2} \left( \sqrt{1 + 4 \frac{\asbt(\gamma)}{(1-\gamma +m)^2}} -1 \right) - \frac{\asbt(\gamma)}{1-\gamma + m}\right], \notag \\
\asbt(\gamma) & = \frac{\asb}{1 + \frac{\asb \beta_0}{4 N_c} \left[\frac{1}{2} \chi_0(\gamma, 0) - \frac{5}{3} + 2(1 + 2 Y/3) \right]},
\end{align}
where $\gamma  = \tfrac{1}{2} + i \nu$ and $Y = 2.343907$, while $\beta_0 = 11 N_c/3 - 2/3 n_f$ with $n_f = 4$ in the following.  The above expression are inspired from expressions used in the  more complete NLO BFKL studies  \cite{Hentschinski:2012kr, Hentschinski:2013id} and should be therefore understood as a first attempt to take into account such effects in an approximate way.
Our numerical results are shown in Fig.~\ref{fig:vanilla_l1},
right. For this study we explore two combinations of photon
virtualities. For red curve we explore a hierarchy of photon virtualities,  $Q_A^2 = 16$~GeV$^2$ and 
$Q_B^2 = 4$~GeV$^2$, which is well placed within the perturbative region of QCD. For the single Pomeron exchange contribution $\sigma_1 \sim e^{\lambda Y}$, we find an approximate Pomeron intercept   of $\lambda \simeq 0.242$, which is in
the range of phenomenological observed intercepts. The uncertainty band corresponds then to  variations of the renormalization scale $\mu \in [Q_A/2, 2Q_A]$, which determines the value of the running coupling constant. In
this case, the Pomeron loop correction amounts to a $8\%$ correction
for $x=10^{-6}$ and to a $5\%$ correction for $x=10^{-5}$. The
relative size of the Pomeron loop contribution is suppressed since the
overall factor of $\asb^4$ is not yet compensated by the growth of the
double Pomeron exchange. The blue curve corresponds on the other hand
to $Q_A^2 = Q_B^2 = 2$~GeV$^2$, which still justifies a perturbative treatment, but which is placed at the boundary to non-perturbative physics. For the  single Pomeron exchange we find in this case an effective intercept of $\lambda \simeq
0.276$. For this second scenario, the Pomeron loop contribution is
significantly enhanced and amounts to a $39\%$ correction at
$x=10^{-6}$ and $24\%$ correction at $x=10^{-5}$.  The result is of
potential interest for the study of $J/\psi$ production in exclusive
vector meson production in photon proton collisions at LHC, where the
hard scale is provided by the charm mass,
$m_c^2 \simeq 1.6-2.0$~GeV$^2$ and one probes hard scales of a similar magnitude.

\section{Conclusions and Outlook}
\label{sec:concl}

In this article we reviewed  existing results on the
determination of the Pomeron loop correction to single Pomeron
exchange within perturbative QCD. As a new result, we provide explicit
expressions which provide a precise mapping of the single Pomeron and
Pomeron loop contribution to the  dipole-dipole amplitude, including the precise relation to the conformal solution of the non-forward BFKL equation by Lipatov.
As already outlined in  \cite{Braun:2013tha}, this relation is needed to evaluate the Pomeron loop, using the projection of the triple Pomeron vertex onto conformal eigenfunctions by Korchemsky \cite{Korchemsky:1997fy}. Our result further uses previously obtained results for the BK equation and dipole amplitude obtained within the high energy effective action, which allows to obtain the Pomeron loop directly from contracting effective reggeized gluon fields. The determination of the remaining integrals over transverse coordinates in Eq.~\eqref{eq:A2f} could however been only carried for the phenomenologically most relevant case of zero conformal spin inside the Pomeron loop. Even though contributions with non-zero conformal spin decrease with increasing energy, it would be desirable to investigate further the relevance of these contributions in the future. 

 While our
results are still limited to conformal spin zero, they allow already
for a first good estimate of the phenomenological relevance of the
Pomeron loop configuration.  Unlike \cite{Braun:2013tha}, we do not
find that the Pomeron loop configuration is expected to dominate
already at LHC energies. We find in general that the size of the Pomeron loop correction depends first of all strongly on the value of strong coupling constant, since the Pomeron loop is proportional to its fourth power. If resummed prescription for the running coupling constant, such as Eq.~\eqref{eq:omega_resummed} are also used for this overall constant, the Pomeron loop would be  naturally strongly suppressed. If on the other hand the strong coupling is directly determined through the external hard scales, the Pomeron loop can  provide a correction of the order of $24\% - 39\%$ for small hard scales and lowest currently accessible values of $x$.

We therefore conclude that  even though  we do not confirm the scenario outlined in \cite{Braun:2013tha}, i.e. that the Pomeron loop configuration exceeds the single Pomeron exchange contribution already at LHC energies, we also cannot discard Pomeron loops  for current phenomenological studies entirely. Future studies should address the determination of the Pomeron loop contribution for the phenomenologically more relevant process of DIS as well as exclusive photoproduction of vector mesons on a proton. Another interesting research task, which we leave for the future, is the determination of the complete resummed 4 reggeized gluon state, which includes correction due to zero, one and two triple Pomeron vertices. We hope to return to this case in the near future. Finally, for the scenarios where the Pomeron loop can provide already a sizeable correction, it would be interesting to explore whether it is sufficient to evaluate the 2 Pomeron loop correction or whether on all order resummation of Pomeron loops needs to be attempted, including possibly an extension to a finite number or colors.

\section*{Acknowledgments}
MH acknowledges support through Secihti project CBF-2026-49 and is grateful to Krzysztof Kutak for useful conversations.

\end{document}